\documentclass[aps,physrev,preprint,superscriptaddress]{revtex4-2}
\usepackage{graphicx} % Required for inserting images
\usepackage{amsmath}

\begin{document}

\title{Enhanced THz third--harmonic generation in graphene via hot--carriers in a topological cavity}

\author{Spyros Doukas}
\email{spdoukas@phys.uoa.gr}
\affiliation{Department of Physics, National and Kapodistrian University of Athens, 15784, Athens, Greece}
\author{Ioannis Katsantonis}
\affiliation{Department of Physics, National and Kapodistrian University of Athens, 15784, Athens, Greece}
\affiliation {Foundation of Research and Technology Hellas, Institute of Electronic Structure and Laser, 71110, Heraklion, Greece}
\author{Thomas Koschny}
\affiliation{Ames National Laboratory and Department of Physics and Astronomy, Iowa State University, 50011, Ames, Iowa, United States}
\author{Elefterios Lidorikis}
\affiliation{Department of Materials Science and Engineering, University of Ioannina, 45110, Ioannina, Greece}
\author{Anna C. Tasolamprou}
\affiliation{Department of Physics, National and Kapodistrian University of Athens, 15784, Athens, Greece}
\affiliation {Foundation of Research and Technology Hellas, Institute of Electronic Structure and Laser, 71110, Heraklion, Greece}

\begin{abstract}
Graphene is a promising material for nonlinear THz applications owing to its high third--order susceptibility and tunable optical properties. Its strong nonlinear response, driven by free--carrier thermodynamics, facilitates efficient third--harmonic generation (THG), which can be further enhanced using resonant metasurfaces and plasmonic structures. In this study, we use a platform that integrates graphene ribbons at the interface of topological photonic crystals designed to boost THG at low THz frequencies. To accurately capture the temperature dependence of the nonlinear optical response,  we perform coupled simulations to model graphene's optical and thermodynamic responses under THz photoexcitation, incorporating photoinduced hot carrier effects within a multiphysics computational framework. This approach unveils the influence of carrier heating on THG, providing a more accurate description of nonlinear THz processes and revealing previously overlooked mechanisms of third--order conductivity modulation. We show that the elevated carrier temperature promotes high conversion efficiencies at moderate input intensities, within a design that operates in both reflection and transmission modes for advanced ultrafast optoelectronic devices in the THz regime. 
\end{abstract}

\maketitle

\section{Introduction}
The terahertz (THz) regime bridges electronic and photonic systems by exploiting the distinctive properties of electromagnetic waves in the 0.1--10 THz range. Located between microwaves and infrared, THz radiation enables a wide range of technological applications, including imaging, spectroscopy, security screening, and wireless communications~\cite{Dhillon_2017,Rappaport201978729,Degl'innocenti20221485}. Nonlinear optical effects in this frequency region have attracted interest, both for investigating fundamental principles of light--matter interactions, as well as for enabling key devices, such as efficient nonlinear THz frequency converters and high--harmonic generation for ultrafast information and communication technologies~\cite{Jiang20242326,Leitenstorfer2023, Koulouklidis2020,Lee201465}. High--harmonic generation requires strong electromagnetic fields interacting with a nonlinear material. In the THz regime, the typical materials used as a high--harmonic generation platform include superconductors and  transition metal oxides~\cite{Yang2019}. Recently, however, graphene, an atomically thin material, has been found to possess some of the highest nonlinear coefficients ever observed in THz, potentially surpassing all known materials. This opens up new avenues for graphene--based applications in ultrafast optoelectronics~\cite{Mics2015,Hafez2020}.

Graphene predominantly exhibits a Drude--like response in the THz regime, which is a result of its easily generated and controlled free carriers~\cite{Ferrari20154598}. Its optical properties can be significantly modulated through chemical doping and/or the application of an external electrostatic or magnetic field~\cite{doukas2018deep,Guo20172989} or in  an ultrafast manner by optical pulse excitation~\cite{doi:10.1021/acsphotonics.8b01595}. 
On top of that, graphene's optical response is nonlinearly dependent on the driving THz field strength. This strong nonlinear response stems from the collective thermodynamic behavior of its free carrier population~\cite{Mics2015,Hafez2020}. Specifically, the THz radiation absorbed via intraband transitions is converted into electronic heat, raising the electronic temperature and modulating its conductivity. Ultimately, the ultrafast conductivity modulation of graphene is defined by the complex interplay between the heating and cooling of its carriers. The former originates from the absorption of intense THz radiation~\cite{Mics2015,Hafez2020} while the latter occurs through carrier--phonon scattering, which transfers the absorbed energy from graphene's thermalized carriers to its lattice~\cite{massicotte2021hot}. This mechanism explains and quantitatively reproduces the recent experimental nonlinear findings, including saturable absorption~\cite{WOS:000345578800018,doi:10.1021/acs.jpclett.4c03138,Hwang2013}, self--induced modulation~\cite{koulouklidis2022ultrafast,Tani2012,PhysRevB.89.041408} and nonlinear wave conversion. Additionally, graphene has been shown to support extremely efficient generation of THz high harmonics at room temperature and under ambient conditions~\cite{Hafez2018,soavi2019hot,AlonsoCalafell2021,doi:10.1021/acsphotonics.3c00543}. Graphene monolayers are centrosymmetric, and therefore the second--order nonlinear response is typically negligible within the dipole approximation; however, it has an exceptionally strong third--order nonlinear susceptibility, with reported values up to $\chi^{(3)}=10^{-9} ~\text{m}^{2}/\text{V}^{2}$~\cite{Glazov2011,PhysRevX.3.021014,Hafez2018}. Third--harmonic generation efficiencies in single--layer graphene on a uniform dielectric substrate have reached values as high as 0.1\% with relatively moderate pump field strengths of THz,  10 to $90$ kV/cm \cite{Hafez2018}. These efficiencies have been shown to improve with the use of electrical gating and the subsequent increase in the density of free carriers~\cite{doi:10.1126/sciadv.abf9809}.
To enhance third--harmonic efficiency conversion of graphene, several schemes have been employed; for example, in Ref.~\cite{Deinert20211145} a grating metasurface with 1\% field conversion efficiency at incident field strengths $30~\text{kV}/\text{cm}$ is demonstrated. Similar performances are reported in Ref.~\cite{DiGaspare2024}, in which split--ring--resonators are used to force local field enhancement. 

To improve THG efficiency, various strategies based on metamaterials,  photonic crystals, photonic cavities,  multi--stacks and, among them, plasmonic systems are investigated \cite{jin2017enhanced, doi:10.1021/acsphotonics.3c00491, maleki2025strategies, theodosi20212d, doi:10.1021/acsphotonics.3c01762, D2NR06286K}, most of them operating in reflection mode. In fact,  graphene has emerged as a highly promising platform for plasmonics in the far--infrared and THz frequency range~\cite{PhysRevB.80.245435,Ju2011,doi:10.1021/acs.nanolett.4c04615}, while traditional plasmonic systems are  based on metals operating at high optical frequencies~\cite{Zhao2011,Tasolamprou2022}. Owing to its lower electron density compared to conventional metals and its two--dimensional nature, graphene exhibits a significantly lower plasma frequency, aligning with the far--infrared and THz spectral regions. Its 2D nature, in addition, enables exceptional field confinement and enhancement, surpassing that of traditional metal--based plasmonics~\cite{PhysRevB.80.245435,Ju2011,doi:10.1021/acs.nanolett.4c04615}. As a single--atom--thick material, graphene also exhibits unique electrical, optical, and thermal properties, making it an ideal candidate for next--generation plasmonic devices. 

An efficient way to manipulate the properties of plasmons, i.e., frequency, spatial confinement, localization characteristics, and improve the relevant electromagnetic functionalities in plasmonic devices is through their hybridization with photonic modes. A category with distinct characteristics is topological photonic modes. As in solid--state materials, which exhibit insulating behavior in the bulk while supporting the transport of electrons along their surface without backscattering, photonic structures can be designed to support unidirectional edge states that are robust against disorder and imperfections~\cite{rider2019topological,PhysRevLett.100.013904,tasolamprou2021chiral,lu2014topological,raghu2008analogs}.   Photonic crystals (PCs), 1D, 2D or 3D periodic structures have been considered as an ideal platform for topological photonics due to engineered photonic bandgaps, surface states, cavities with high field localization and advanced light manipulation capabilities~\cite{mavidis2020local,tasolamprou2015frequency,Katsantonis2023}. In the simple configuration of the 1D photonic crystal, the topological invariance is connected to the Zak phase, i.e., the geometric phase acquired by a Bloch wavefunction over the Brillouin zone,  and  distinguishes between different bulk band structures, even in systems that are otherwise spectrally similar~\cite{PhysRevX.4.021017, PhysRevB.93.041415, SHARMA2023113508}. Beyond their importance in fundamental physics, topological systems offer surface states that appear within specific photonic bandgaps and are characterized by increased confinement  and most importantly by the robustness coming from the topological protection, i.e., they are immune to possible imperfections and deformations.

Motivated by these recent advances, this work investigates a platform for enhancing THG in the low--terahertz frequency range. Specifically, we integrate graphene, in the form of a periodic array of micro--ribbons, at the interface between two one--dimensional PCs. The two PCs are engineered to exhibit opposite topological properties, characterized by distinct Zak phases. A topological edge mode is established at the fundamental frequency (FF). Simultaneously, the width and periodicity of the graphene ribbons are tailored to support plasmonic modes that resonate with this edge mode. In addition, a Fabry--Pérot (FP) mode is engineered to appear at the third--harmonic (TH) frequency, collectively resulting in a significant enhancement of graphene's nonlinear optical response. To model the THG process, we incorporate both the linear and third--order nonlinear surface conductivities of graphene into COMSOL Multiphysics finite element method (FEM) solver. Hot--carrier effects--previously overlooked in THz THG modeling--are considered by self--consistently accounting for the elevated electronic temperature induced by THz photoexcitation. This is implemented using a well--established multiphysics simulation framework tailored for graphene~\cite{koulouklidis2022ultrafast, doukas2022thermionic, doukas2022electrical}, integrated within the full--wave FEM environment. We incorporate realistic material parameters and explore a broad range of conditions to assess the third--harmonic power outflow and the overall conversion efficiency.

Our optimized structures achieve a TH conversion efficiency of $-18.4$~dB, corresponding to a power outflow of 2.2 kW/cm, when excited with an intensity of $I_0=$ 150 kW/cm$^2$ at the FF of 2.53 THz. The generated TH waves can be simultaneously extracted in both the forward (transmitted) and backward (reflected) directions. Importantly, we emphasize the role of finite--temperature effects in the simulation framework and demonstrate that elevated electronic temperatures can enhance the THG response. In contrast, assuming carriers in graphene remain at ambient temperature leads to a lower conversion efficiency of $-23.3$~dB and a reduced output power of 0.7 kW/cm$^2$.

The paper is organized as follows: In Section~\ref{topo_cavity} we present the design and aspects of the topological cavity into which the graphene ribbons are to be integrated, in Section~\ref{thermal_response} we present the thermodynamic approach modeling of graphene's response upon THz photoexcitation, while in Section~\ref{sim_procedure} we describe the THG calculation procedure upon incorporating our self--consistent model with the FEM solver. Finally, in Section~\ref{results} we present the optimization process that we follow to enhance THG and discuss our findings.
\section{Methodology}
\subsection{The topological photonic cavity}\label{topo_cavity}
We begin our analysis by designing a planar heterostructure, i.e., the topological photonic cavity, that supports simultaneous resonances at both the FF and the TH. Resonance at the fundamental frequency leads to strong electromagnetic fields that modulate the  graphene plasmons in the hybrid system. At the same time, graphene plasmons allow us to control the resonant frequency while further enhancing the absorption of THz energy. Additionally, resonance at the TH ensures an increased optical density of states, which enhances the radiation emitted from the structure at this frequency, further boosting the efficiency of the THG process.

\begin{figure}[ht]\centering
	{\includegraphics[]{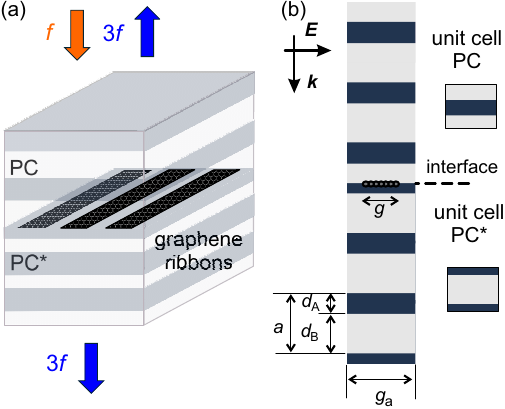}}
	\caption{Schematic illustration of the structure and principle of operation. (a) The structure consists of graphene ribbons embedded in a topological cavity. At the FF frequency of operation ($f$) the structure generates TH waves (3$f$) emitted in both forward (transmission) and backward (reflection) directions. (b) The topological cavity if formed  by interfacing two PCs, PC and PC*, each one containing three unit cells. The unit cells are defined at different inversion centers, as depicted in the corresponding insets. The unit cell size of both PCs is $a$ and it consists of two layers, $d_A$ and $d_B$, of different dielectrics with indices $n_A$ and $n_B$ respectively. The graphene ribbons are incorporated at the interface plane of the two PCs. Their width is $g$ and the distance between the adjacent ribbons' centers is $g_a$. The filling factor of the graphene ribbons is defined as $L=g/g_a$. }
	\label{figPC}
\end{figure}

The structure is illustrated in Fig.\ref{figPC}. It  consists of two  photonic crystals, PC and PC*, each constructed from the same sequence of alternating dielectric layers but with a shifted origin of the unit cell, as shown Fig.\ref{figPC}. Each semi--space consists of three unit cells of type PC and PC*, respectively. The total thickness of the finite structure is equal to six unit cells. In this initial phase of the study we focus  on establishing the photonic cavity and thus omit the graphene ribbons to isolate the purely photonic behavior of the system. The discussion of the plasmonic--photonic coupling, that arises from incorporating graphene ribbons, is elaborated in Appendix \ref{AppA}. 

Each unit cell of the PCs is formed by the same sequence of alternating, optically different layers of high--index material A and low--index material B. For materials A and B we select high--resistivity float--zone silicon (HRFZ--Si) and Polymethylpentene (TPX), respectively. At the frequencies considered here, the refractive indices of the selected materials are \(n_A=3.416\) for HRFZ--Si~\cite{dai2004terahertz} and \(n_B=1.46\) for TPX~\cite{bichon2022complex}, respectively. The materials above were chosen because they present almost no dispersion and negligible losses in the frequency regime we aim to study in this work~\cite{,dai2004terahertz, bichon2022complex}. Their thicknesses are selected as \(d_A=a/6\) and \(d_B=5a/6\), respectively, where \(a=30 \, \mu m\) is the thickness of the unit cells of the two PCs. The layer configurations and dimensional details are illustrated in Fig.~\ref{figPC}(b). Photonic crystal PC starts and ends with the center of layer A, while photonic crystal PC* starts and ends with the center of layer B. This configuration allows us to access a topological  state confined to the PC/PC* interface \cite{PhysRevX.4.021017, PhysRevB.93.041415}.

\begin{figure}[ht]\centering
	{\includegraphics[width=0.95\textwidth]{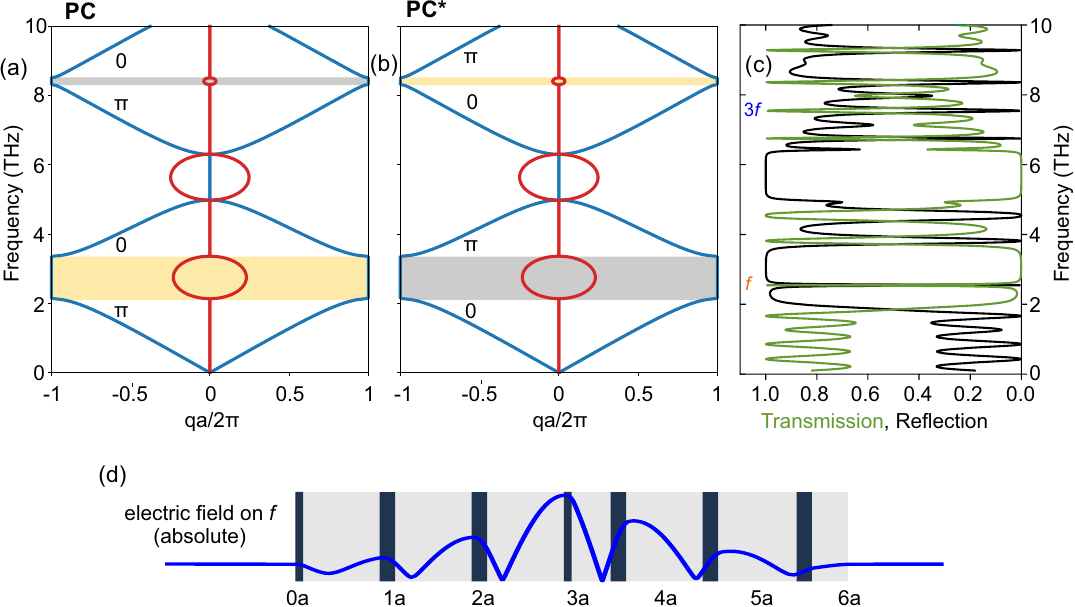}}
	\caption{The band structure of the infinite PC (a) and the infinite PC* (b) with parameters  \(n_A=3.416\) and \(d_A=a/6\), and  \(n_B=1.46\) and \(d_B=5a/6\), where \(a=30 \, \mu m\). For each band the Zak phase is calculated and marked in the panels. (c) Transmission (green curve) and reflection (black curve) for a multilayer consisting of three PC unit cells and three PC* unit cells. At the interface of the PC/PC* regions a topological surface state emerges which lies within the first bandgap, at $f=2.55$~THz.  A Fabry--Pérot mode appears at $3f = 7.65$ which will host the generated third--harmonic waves. (d) Distribution of the absolute value of the electric field of the topological mode in the PC/PC* structure at $f=2.55$~THz. A maximum peak appears at the interface of the two regions.  }
	\label{fig_bandstructure}
\end{figure}

The two photonic crystals, PC and PC*, have identical bandstructure [see Fig.~\ref{fig_bandstructure}(a)]. The topological state emerges at the interface between the two photonic crystals, PC and PC*, within specific bandgaps. The presence or absence of these interface states is governed by the bulk properties of the corresponding infinite photonic crystals and originates from the fact that the origins of their unit cells are defined at different inversion centers. This shift in the origins of the unit cells results in different topological characteristics between the two PCs, which can be quantified through the geometric phases of the bulk bands, known as Zak phases~\cite{PhysRevX.4.021017,ZakPhysRevLett.62.2747,SHARMA2023113508}. The Zak phase represents the phase accumulated by an eigenvector as it travels across the entire Brillouin zone of a periodic structure. When two PCs with different Zak phases are placed next to each other, the change in topological character at the interface leads to the formation of localized modes--called topological interface states--within the bandgap. In contrast, if the Zak phases of the two PCs are the same, such topological states are typically absent. 

To investigate the topological properties of our system, we calculate the Zak phase of both bands, in each infinite bulk photonic crystal PC and PC*~\cite{PhysRevX.4.021017}. The Zak phase in this 1D Bloch band system with Bloch eigenstates is defined as:  

\begin{equation}
	\theta^{\text{Zak}} = i \int_{\text{BZ}} \langle u_k | \partial_k u_k \rangle \, dk
\end{equation} 
where the integration is over the Brillouin zone (BZ),  \(u_k\) is the cell--periodic part of the Bloch wavefunction, and \(\partial_k\) denotes the derivative with respect to the Bloch momentum \(k\). In the numerical study, we perform full--wave eigenvalue analysis and we calculate the discretised Zak phases from
\begin{equation}
	\theta_m^{\text{Zak}} = - \sum_{j=0}^{N-1} \mathrm{Im} \left[ \log \left( \langle u_{m,k_j} \mid u_{m,k_{j+1}} \rangle \right) \right]
\end{equation} 
In the above, \(u_{m,k}(z)\) is the periodic--in--cell part of the Bloch electric field of a state on the $m$--th band with wavevector k, i.e., \(u_{m,k}(z)=E_{m,k}(z)e^{ikz}\) with \(E_{m,k}(z)e^{ikz}\) being the $m$--th eigenstate of the system. The  overlap between the neighboring Bloch eigenfunctions is given by \(
\langle u_{m,K_j} \mid u_{m,K_{j+1}} \rangle = \int_{\text{unit cell}} \varepsilon(z) \, u_{m,K_j}^*(z) \, u_{m,K_{j+1}}(z) \, dz
\), where  \(\epsilon(z)\) is the permittivity along the unit cell. The Zak phase \(\theta_m^{Zak}\) and topological invariants are obtained for each  band  and are shown in Fig 2 (a)--(b). Having calculated the  Zak phases, \(\theta_m^{Zak}\), and since there are no band crossings in our system,  the topological invariant is defined as 

\begin{equation} \label{eq:2}
	\text{sgn}[s^{(n)}] = [(-1)^n  \exp (i \sum_{m=0}^{m=n-1} \theta_m^{Zak})]
\end{equation} 
where $n$ counts the bandgaps and $m$ counts the bands.
Note that in the calculation of topologically invariant Eq.~\ref{eq:2} of each $n$ gap, we count all the $m$ bands below this gap. For the topological state to emerge,  the sign of $s^{(n)}$ should be different in each photonic crystal PC and PC*.

We summarize all the above in  Fig.~\ref{fig_bandstructure} where we present the analysis of the topological photonic cavity formed by PC and PC*. In panels Fig.~\ref{fig_bandstructure}(a) and Fig.~\ref{fig_bandstructure}(b) we present the band structures of the infinite PC and PC*, respectively, along with the Zak phases for each band. Given that the two PCs only differ in the origin of their unit cell definition, their bandstructure and, hence, their bandgaps are the same. However, a change in the topological phase is caused by the shifted definition of the unit cell in PC and PC*, and guarantees a robust topologically protected surface state at the interface between PC and PC*. We take initially PC, Fig.~\ref{fig_bandstructure}(a), and the first bandgap. The Zak phase of the band below the bandgap is $\pi$. 
Thus,  we calculate a positive sign, $\text{sgn}[s_{\text{PC}}^{(1)}]=+1$. We mark the first bandgap, that corresponds to the positive sign of \(s^{(1)}\), with yellow. For the same gap, in the PC* crystal, we calculate a negative sign $\text{sgn}[s_{\text{PC*}}^{(1)}]=-1$.  We mark the bandgap that corresponds to the negative sign with grey shadowing. Due to the different signs for PC and PC*, the bandgap can host a topological surface state. 
Similarly, for the second bandgap, in PC we get $\text{sgn}[s_{\text{PC}}^{(2)}]=-1$ and for PC* $\text{sgn}[s_{PC*}^{(2)}]=-1$, that is negative for both  PC and PC*. Hence, the second bandgap cannot host a topological surface state. Finally, for the third bandgap we get for PC  $\text{sgn}[s_{PC}^{(3)}]=-1$  and $\text{sgn}[s_{PC*}^{(3)}]=+1$. 

The above procedure is the recipe for  designing a simple multistack topological photonic structure, with a surface  state appearing at the first gap. The surface state manifests itself as a peak in the scattering properties of the structure. In Fig.\ref{fig_bandstructure}(c) we present the transmission and reflection of the structure. We observe a full transmission resonance, at frequency 2.55 THz, which is related with the topological surface state. This topological surface state leads to a strong electric field enhancement at the interface of PC/PC*. This is presented in Fig.~\ref{fig_bandstructure}(d) where we plot the absolute value of the electric field calculated by eigenvalue analysis in a finite system. This serves as the FF resonance. At the same time, we observe another transmission peak at 7.65 THz which serves as the TH resonance. This resonance is formed by bulk propagating modes in a FP configuration. The fact that the TH mode propagates in the photonic crystal ensures that the converted energy will radiate outside of the cavity, producing outgoing waves in both the forward and backward directions. 

\subsection{Graphene's thermodynamic response}\label{thermal_response}
The linear and nonlinear response of graphene to THz photoexcitation is fundamentally governed by the collective thermodynamic behavior of its free carriers \cite{mics2015thermodynamic, Hafez2018, Hafez2020, massicotte2021hot}. This response is primarily driven by the absorption of THz radiation through intraband transitions \cite{Hafez2018, massicotte2021hot}, coupled with the rapid internal thermalization of free carriers. Upon light absorption, graphene carriers redistribute their acquired energy across the entire carrier population--due to electron--electron interactions--\cite{massicotte2021hot}, leading to the establishment of a common electronic temperature $T_e$ within an ultrafast timescale in the order of tens of fs \cite{massicotte2021hot, tomadin2013nonequilibrium}. Concurrently, energy dissipation from the electronic system to the phonon bath occurs on a comparatively slower timescale in the ps regime \cite{massicotte2021hot, tomadin2013nonequilibrium}, thereby sustaining efficient carrier heating. The initial graphene carrier density, at $T_e = 300$~K, is dictated by its Fermi level $E_F$, which is determined by fabrication and transfer processes of graphene on the selected substrate and is tunable via electrostatic gating \cite{doukas2018deep}. The photoinduced increase in electronic temperature, $T_e$, broadens the Fermi--Dirac distribution \cite{massicotte2021hot, koulouklidis2022ultrafast} and reduces the chemical potential $\mu_{\text{SLG}}$ \cite{koulouklidis2022ultrafast, doukas2022thermionic}. This occurs due to the conservation of net charge density \cite{mics2015thermodynamic, koulouklidis2022ultrafast}, expressed as $n_e(\mu_{\text{SLG}}, T_e) - n_h(\mu_{\text{SLG}}, T_e) = $ const., where $n_e(\mu_{\text{SLG}}, T_e)$ and $n_h(\mu_{\text{SLG}}, T_e)$ are the electron and hole densities in graphene, respectively \cite{doukas2022thermionic}. For sufficiently doped graphene, where $E_F>k_BT_e$ (with $k_B$  being Boltzmann's constant), an increase in $T_e$ leads to a reduction in its linear intraband conductivity  $\sigma^{(1)}$ \cite{massicotte2021hot, mics2015thermodynamic, kovalev2021electrical}.    

\begin{figure}[ht]\centering
	{\includegraphics[width = 1.0\textwidth]{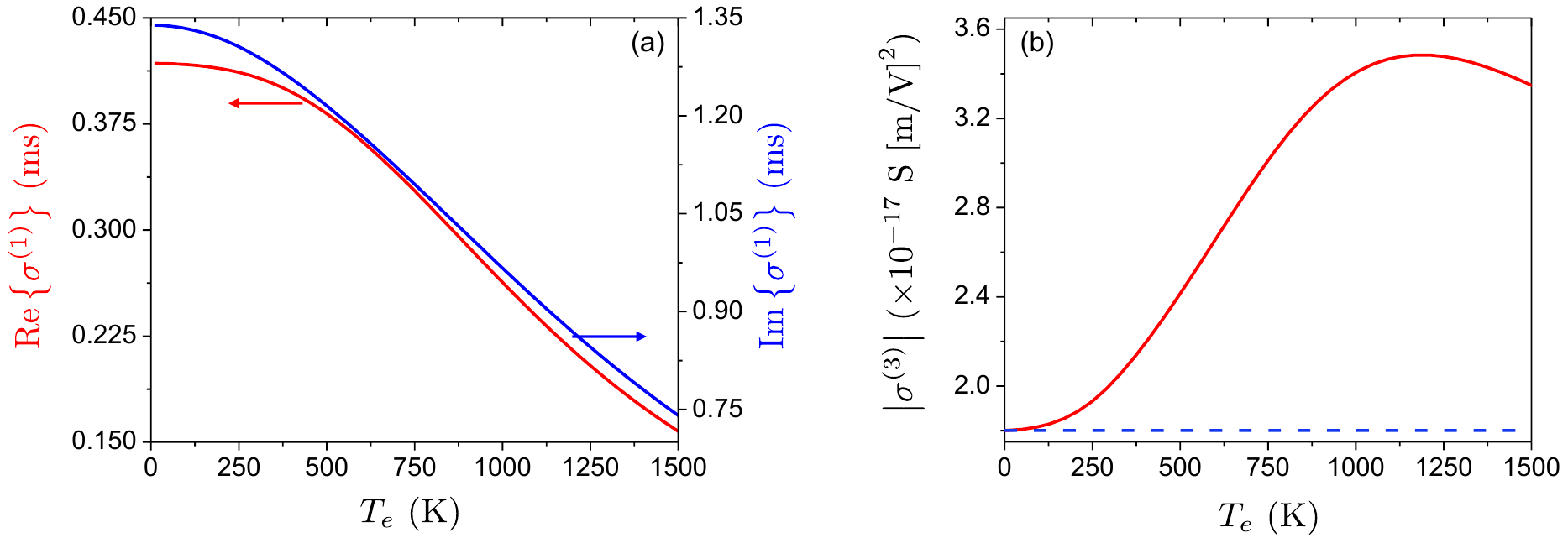}}
	\caption{(a) Real (left) and imaginary (right) part of graphene's linear intraband conductivity $\sigma^{(1)}$ at 2.55 THz as a function of carrier temperature $T_e$ assuming $E_F$=0.2 eV and linear energy dependence of momentum relaxation time with a proportionality constant $\gamma = 1000$  fs/eV. (b) Graphene's third--order nonlinear conductivity for $E_F=$ 0.2 eV and fundamental frequency 2.55 THz. The dotted line plots the result at constant temperature $T_e = 0$ as for Eq. \ref{cond_Te_0}.}
\label{fig_cond}
\end{figure}

The frequency dependent linear intraband conductivity of single--layer graphene (SLG), which dominates its response in THz frequencies, is given by \cite{koulouklidis2022ultrafast, doukas2022thermionic}:
\begin{equation}
	\begin{gathered}
		\sigma^{(1)}(\omega, T_e, \mu_{\text{SLG}}) = \frac{ie^2}{\pi \hbar^2} \times  \\[8pt] \int_0^\infty \mathrm{d}E \frac{E}{\omega + i\tau^{-1}(E)} \left[ \frac{\partial f_{\text{FD}}(-E; \mu_{\text{SLG}}, T_e)}{\partial E} - \frac{\partial f_{\text{FD}}(E; \mu_{\text{SLG}}, T_e)}{\partial E} \right]
	\end{gathered}
	\label{cond_eq}
\end{equation}
with $f_{\text{FD}}(E; \mu_{\text{SLG}}, T_e)$ the Fermi--Dirac distribution and $\tau(\epsilon)$ the energy--dependent momentum scattering time \cite{koulouklidis2022ultrafast, mics2015thermodynamic}. The latter is calculated assuming long--range scattering with Coulomb impurities, that is $\tau(\epsilon) = \gamma \epsilon$ \cite{das2011electronic}, which is the dominating mechanism for CVD grown graphene supported by dielectric substrates \cite{koulouklidis2022ultrafast, mics2015thermodynamic, Hafez2020}. This description of carrier relaxation has been utilized in our previous study, yielding perfect agreement with experimental measurements in THz photoexcited graphene \cite{koulouklidis2022ultrafast}. Throughout this work, we assume a constant graphene Fermi level of $E_F = 0.2$ eV. This doping level is typically achievable in chemical vapor deposition (CVD)--grown graphene due to fabrication and transfer processes \cite{goniszewski2016correlation}, thereby eliminating the need for additional electrostatic gating. Avoiding the complexities associated with gating is advantageous for practical implementation, as it simplifies both the fabrication and integration of the proposed configuration. Moreover, we assume a proportionality constant for Coulomb--scattering \( \gamma = 1000 \) fs/eV, which results in a moderate momentum relaxation time of \( \tau = 0.2 \) ps at room temperature. Figure \ref{fig_cond}a shows the intraband conductivity \( \sigma^{(1)} \) at 2.55 THz as a function of \( T_e \), illustrating its decline with increasing $T_e$.

The third--order nonlinear surface conductivity of graphene at the $T_e = 0$ limit is given by \cite{cheng2014third, mikhailov2016quantum}: 

\begin{equation}\label{cond_Te_0}
\sigma^{(3)}(\omega, E_F, 0) = i \sigma_0^{(3)} \bar{\sigma}^{(3)}(\omega, E_F, 0)
\end{equation}
with $\sigma_0^{(3)} = \dfrac{N_f e^4 \hbar v_F^2}{32\pi}$, $N_f=4$, $v_F = 10^6$ m/s the Fermi velocity in graphene and the term $	\bar{\sigma}^{(3)}$ given by :
\begin{equation}
\bar{\sigma}^{(3)}(\omega, E_F, 0) = \frac{17G(2\left|E_F\right|, \hbar \omega) - 64G(2\left|E_F\right|, 2\hbar \omega) + 45G(2\left|E_F\right|, 3\hbar \omega)}{24(\hbar \omega)^4}
\end{equation} 
with:
\begin{equation}\label{G_eq}
G(x,y)= \text{ln}\left| \dfrac{x+y}{x-y}\right|+i \pi \Theta\left(y-x\right).
\end{equation} 
In the above relationship $\Theta(y-x)$ is the Heaviside step function equal to 1 for $y\geq x$ and 0 for $y<x$. 

Given the analytical expression for the third--order conductivity in the $T_e = 0$ limit, the effects of finite electron temperature can be incorporated using the Maldague identity \cite{giuliani2008quantum}. To achieve this, one may use Eq.\ref{cond_Te_0} with the substitution $E_F \rightarrow \epsilon$ and integrate over the entire energy spectrum \cite{soavi2019hot, rostami2016theory, cheng2015third, ghaebi2024ultrafast}. Doing so, the $T_e$ dependent expression becomes: 
\begin{equation}\label{cond_Te_finite}
\sigma^{(3)}(\omega, E_F, T_e) = \dfrac{1}{4 k_B T_e} \int_{-\infty}^{\infty} d\epsilon \, \dfrac{\sigma^{(3)}(\omega, \epsilon, 0)}{\cosh^2 \left( \dfrac{\epsilon - \mu_{\text{SLG}}}{2 k_B T_e} \right)}
\end{equation}
Figure \ref{fig_cond}b plots the $T_e$--dependent third--order graphene conductivity $\sigma^{(3)}(\omega, \mu_{\text{SLG}}, T_e)$  at 2.55 THz (solid line), assuming graphene parameters as discussed above. In the same plot we show the corresponding results at the constant $T_e$ = 0 limit, i.e, Eq. \ref{cond_Te_0} (dashed line). As expected, the relationships Eqs. \ref{cond_Te_0} and \ref{cond_Te_finite} match to the same value at $T_e$ = 0, yielding a third--order susceptibility for graphene $\chi^{(3)}=-i\sigma^{(3)}/\omega \epsilon_0 d_{\text{SLG}} \approx 3.8^{-10}$ m$^2$\,V$^{-2}$ \cite{rostami2016theory, Hafez2018}. The inclusion of $T_e$ effects, however, modulates the value of $\sigma^{(3)}$, presenting a peak value at $T_e \approx 1100$ K. 
\begin{figure}[ht]\centering
{\includegraphics[width = 1.0\textwidth]{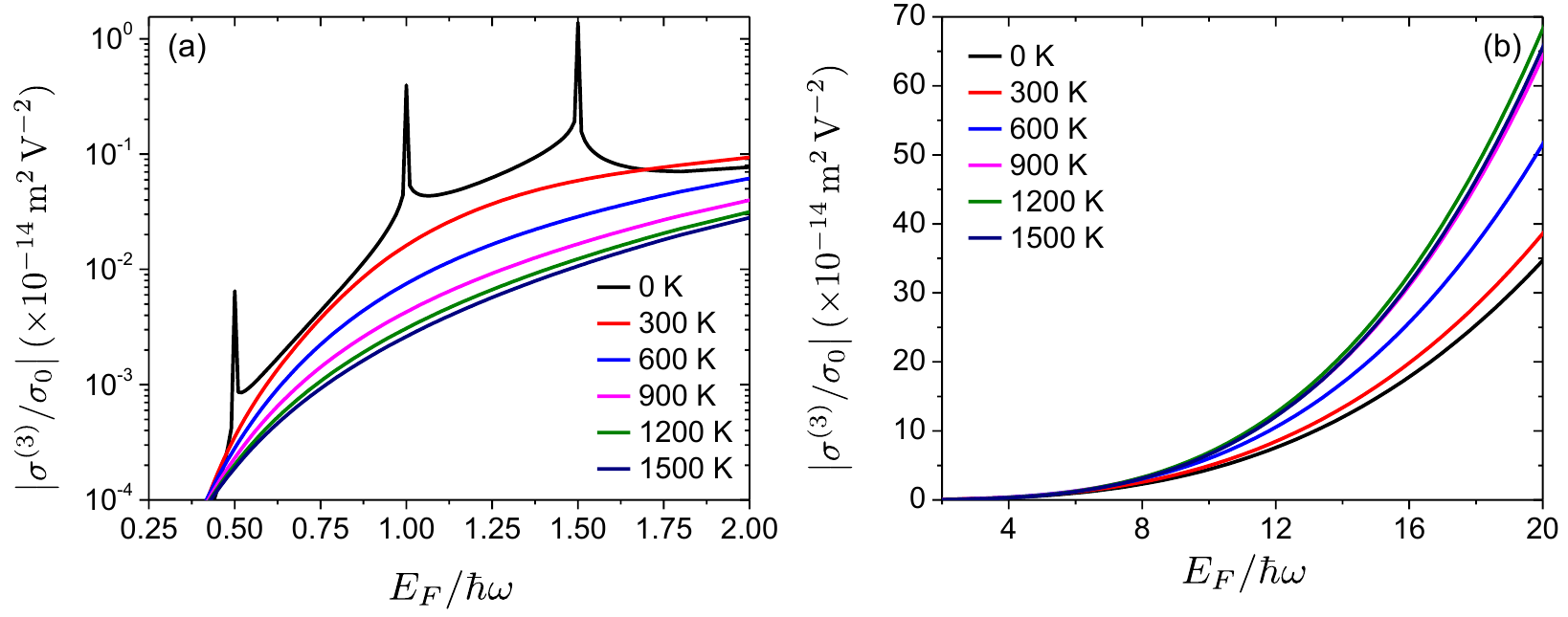}}
\caption{Magnitude of graphene's third--order nonlinear conductivity, normalized to $\sigma_0 = e^2/4\hbar$, as a function of $E_F/\hbar \omega$ ratio for different $T_e$ values. (a) Corresponds to the IR region and (b) to the THz regime. All calculations were performed assuming $E_F$ = 0.2 eV.}
\label{sigma3_vs_en_all}
\end{figure}

It is important to emphasize that the non--monotonic dependence of $\sigma^{(3)}$ on $T_e$ is characteristic of the THz regime. This behavior stems from the structure of $\sigma^{(3)}$ at zero electron temperature, given by Eq. \ref{cond_Te_0}, which features divergent peaks at energies $m \hbar \omega = 2 |\epsilon|$ (with $m = 1, 2, 3$), corresponding to resonant electronic transitions \cite{cheng2014third}. When finite $T_e$ is introduced via Eq. \eqref{cond_Te_finite}, these sharp features are convoluted with a thermal broadening kernel centered at $\epsilon = \mu_{\text{SLG}}$ and with characteristic width $\sim 2k_BT_e$. The resulting integral is highly sensitive to the frequency regime due to this convolution.

In the infrared (IR), where $\hbar\omega$ is comparable to typical graphene Fermi levels (0.2--0.8 eV), $\sigma^{(3)}$ at $T_e = 0$ displays distinct logarithmic peaks at $m \hbar \omega = 2|E_F|$, as described by Eq. \ref{cond_Te_0}. As $T_e$ increases, these features are increasingly smeared out, leading to a monotonic suppression of the nonlinear response \cite{soavi2018broadband} (Fig. \ref{sigma3_vs_en_all}a).

\begin{figure}[ht]\centering
{\includegraphics[width = 1.0\textwidth]{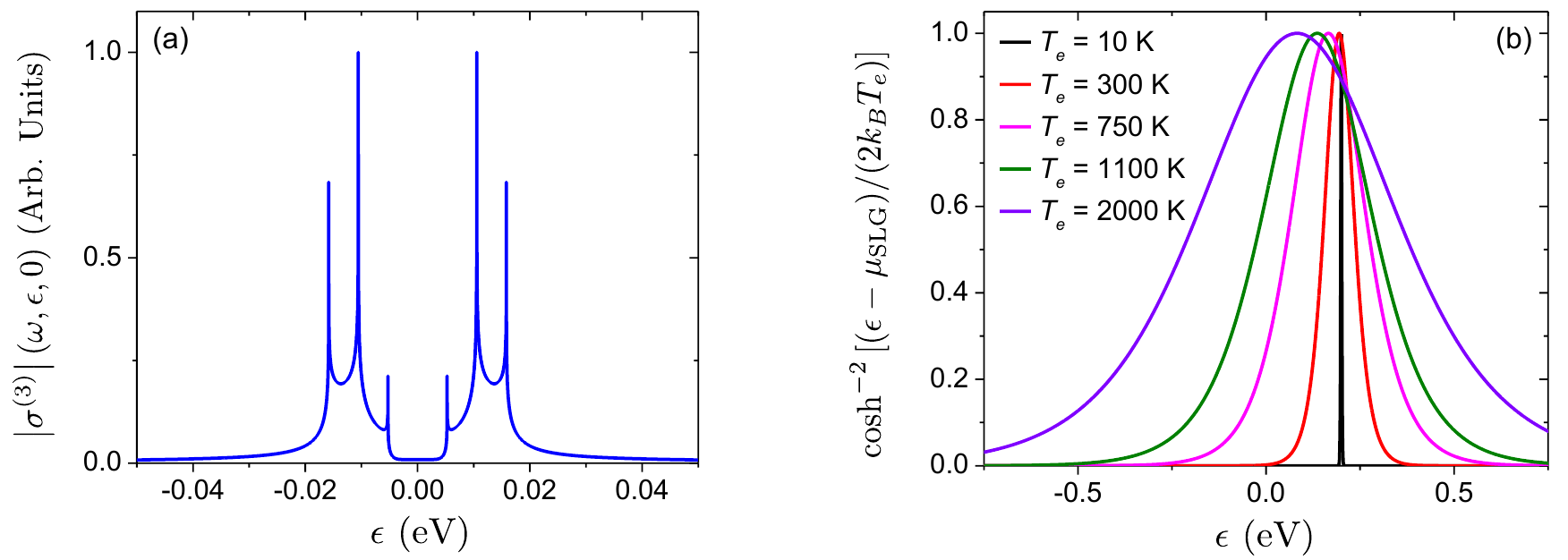}}
\caption{Frequency dependent (a) and temperature dependent (b) integrands of graphene's third--order conductivity (Eq. \ref{cond_Te_finite}) as a function of carrier energy $\epsilon$ in eV. In (a) we assumed FF at 2.55 THz while panel (b) plots the results for different $T_e$ values, highlighting the $\mu_{\text{SLG}}$ reduction and carrier distribution broadening at elevated carrier temperatures.}
\label{sigma3_vs_en}
\end{figure}

\begin{figure}[ht]\centering
{\includegraphics[width = 0.55\textwidth]{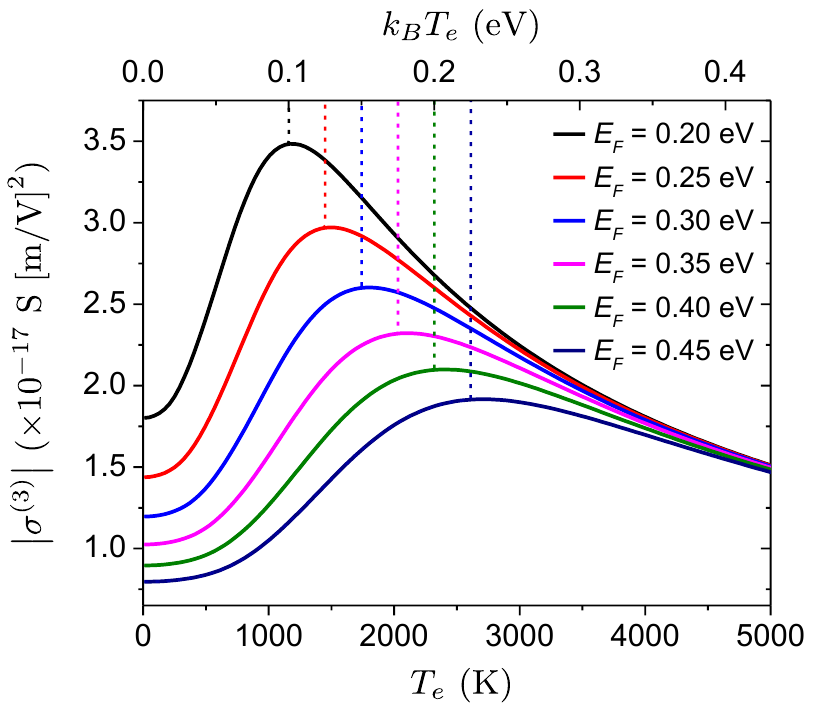}}
\caption{Third--order graphene conductivity, assuming FF at 2.55 THz and different $E_F$ values, as a function of carrier temperature $T_e$. Top horizontal axis denotes the corresponding carrier thermal energy in eV. The dotted vertical lines indicate the regions corresponding to $T_e \approx 2E_F/k_B$.  }
\label{sigma3_vs_Ef}
\end{figure}

Conversely, in the THz regime, with photon energies an order of magnitude lower than typical graphene $E_F$ values, the thermal broadening initially enhances $\sigma^{(3)}$. This can be understood from Fig. \ref{sigma3_vs_en}, which shows the integrand structure of Eq. \ref{cond_Te_finite}. Specifically, Fig. \ref{sigma3_vs_en}a presents the normalized magnitude of $\sigma^{(3)}(\omega, \epsilon, 0)$--for excitation at 2.55 THz--highlighting the narrow resonances at zero temperature. As $T_e$ increases, the thermal weighting function $\text{cosh}^{-2}[(\epsilon - \mu_{\text{SLG}})/(2k_BT_e)]$ broadens (Fig. \ref{sigma3_vs_en}b), allowing the integral to sample these resonances more effectively, thereby amplifying the nonlinear response. However, beyond a critical temperature, the chemical potential $\mu_{\text{SLG}}(T_e)$ shifts downward and the broadening becomes too extended, causing the integral to include off--resonant regions where $\sigma^{(3)}(\omega, \epsilon, 0)$ is diminished. This leads to a maximum of $\sigma^{(3)}$ at a certain $T_e$, a behavior consistently observed for different values of $E_F$ (Fig. \ref{sigma3_vs_Ef}a). It turns out that the optimal $T_e$ as a function of $E_F$, corresponding to the alignment of thermal broadening with the resonance features of  $\sigma^{(3)}(\omega, \epsilon, 0)$, is approximately given by $T_e \approx {E_F} / 2k_B$ (Fig. \ref{sigma3_vs_Ef}a). At even higher $T_e$, above a few thousand K, a large fraction of graphene carriers are thermalized and $\sigma^{(3)}$ converges to a single value, irrespective of the initial $E_F$.

To accurately capture the photoinduced thermal effects, one needs to self--consistently employ a two--temperature model that treats electrons and phonons separately, accounting for their distinct thermalization and cooling timescales \cite{doukas2022thermionic, koulouklidis2022ultrafast, doukas2022electrical}. This approach provides a detailed description of energy dissipation in graphene under continuous--wave (CW) excitation, including electron--phonon scattering both in the optical phonon--branch, through Boltzmann collision integral, and acoustic phonon scattering via short--range disorder assisted supercollisions \cite{doukas2022thermionic}. In this work, we focus on moderate illumination intensities, which lead to electron temperatures in the order of $T_e \lesssim 800 $ K. Due to the significantly larger lattice heat capacity of graphene--approximately three orders of magnitude greater than its electronic heat capacity \cite{doukas2022thermionic, pop2012thermal}--the excess energy from photoexcitation primarily affects the electron system, while the lattice temperature $T_l$ experiences only a modest increase of about $\sim 20$ K. Given this relatively small temperature rise we assume, for simplicity, that the lattice remains at a fixed temperature of $T_l = 300$ K throughout our analysis. We  calculate the $T_e$ rise under CW THz illumination by solving the equilibrium equation \cite{doukas2022electrical,doukas2022thermionic}:
\begin{equation}\label{Te_cw}
\alpha_{\text{SLG}} I_0 = J_{\text{e--ph}} 
\end{equation}
where $\alpha_{\text{SLG}} \equiv \alpha_{\text{SLG}}(E_F, T_e)$ is the absorbance of graphene in the given configuration, $I_0$ is the incident power density, and $J_{\text{e--ph}} \equiv J_{\text{e--ph}}(E_F, T_e)$ the thermal current density dissipated into the phonon bath via electron--phonon interactions.  The term $J_{\text{e--ph}}$ accounts for carrier--phonon interactions, explicitly incorporating contributions from both optical phonon branches at the K--point and the doubly degenerate $\Gamma$--point of graphene's Brillouin zone \cite{ferrari2006raman, ferrari2007raman}. Additionally, our analysis includes cooling via disorder--assisted supercollisions \cite{song2015energy, song2015energy}, assuming short--range scatterers with a mean free path of $l=$ 100 nm on par with our previous works \cite{doukas2022thermionic, koulouklidis2022ultrafast}. 

\subsection{Simulation approach for the calculation of third--harmonic generation}\label{sim_procedure}
The proposed configuration consists of continuous SLG ribbons of width $g$, arranged in a periodic array with period $g_a$, at the interface between two PCs, as shown in Fig. \ref{figPC}. The linear and nonlinear optical responses of these structures are acquired using the FEM in the frequency domain via the COMSOL Multiphysics solver. We assume SLG micro--ribbons significantly longer than their width, ensuring uniformity along the z--axis, and perform 2D simulations to compute their optical response. Periodic boundary conditions are applied along the x--direction to model the periodic array, with only a single period included in the simulation domain. The structure is illuminated under normal incidence at the FF by a TM--polarized ($\mathbf{E} \equiv \mathbf{E_x}$) plane wave propagating along the y--axis (Fig. \ref{figPC}b).

Graphene is incorporated into the simulations as a surface boundary, with its linear and nonlinear surface conductivities defined by Eqs. \ref{cond_eq} and \ref{cond_Te_finite}, respectively. During the third harmonic generation (THG) process, three photons with angular frequency $\omega_{FF}$ are annihilated to produce a single photon at the third harmonic (TH) frequency, with angular frequency $\omega_{TH}=3 \omega_{FF}$. Consequently, a nonlinear surface current density is induced in graphene, expressed as:
\begin{equation}
	\boldsymbol{J}_{\omega_{TH}} = \sigma^{(1)}(\omega_{TH},T_e, \mu_{\text{SLG}})\boldsymbol{E}(\omega_{TH})+	\sigma^{(3)}(\omega_{TH},T_e, \mu_{\text{SLG}})\boldsymbol{E}^3(\omega_{FF})
\end{equation}
The TH waves, generated in graphene, are radiating from the interface of the PCs in both directions. We employ a perturbative approach, in which the THG effects are weak enough so that they do not significantly affect the fundamental field. In this approximation, the THG process can be simulated by decoupling the nonlinear problem into two separate linear frequency simulations in the full--wave FEM solver: one at the FF and another at the TH frequency \cite{jin2017enhanced}. The investigated configuration is first excited at the FF, and its response is computed using the methods outlined previously. A second simulation is then performed without external wave excitation, where the induced nonlinear current density acts as a source radiating at the TH frequency. A sufficiently fine mesh is incorporated in all FEM simulations to accurately resolve the TH waves. The radiated TH power outflow is determined by computing the Poynting vector through the simulation boundaries. This is performed for both forward (transmitted) and backward (reflected) propagating TH waves. From this, we extract the conversion efficiency (CE) of the THG process, given by $\text{CE}_{i} = P_{i, TH}/P_{in,FF}$ where the index $i$ denotes the TH power of either transmitted or reflected waves, and $P_{in,FF}$ the incident FF power.

We self--consistently calculate the electrical (chemical potential, carrier density), optical (linear and nonlinear surface conductivities), and thermal (electron--phonon thermal current density) properties of graphene. Equation \ref{Te_cw} is then used to relate the incident FF power density, $I_0$, to the resulting increase in electron temperature, $T_e$, while accounting for $T_e$-dependent linear and nonlinear optical responses as discussed above.

\section{Results and discussion}\label{results}
\begin{figure}[ht]\centering
	{\includegraphics[width = 1.0\textwidth]{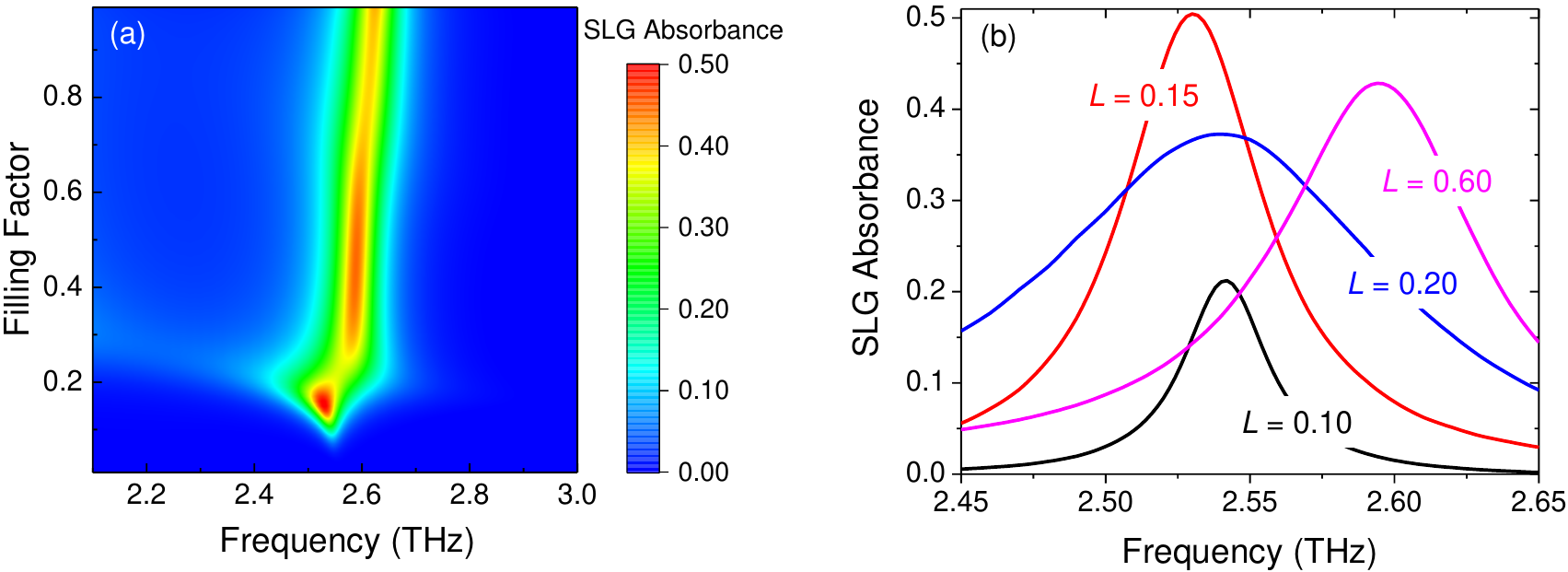}}
	\caption{(a) SLG absorbance colormap as a function of FF frequency and filling factor $L = g/g_a$. (b) SLG absorbance spectra for representative $L$ values in the vicinity and far from resonant conditions.}
	\label{fig_abs_linear}
\end{figure}
We begin our investigation by varying the geometric characteristics of the SLG ribbons to excite plasmonic modes that resonate with the topological surface state of the PCs. Specifically, we fix the ribbon periodicity at $g_a=8 \,\mu$m and vary the ribbon width $g$. Figure~\ref{fig_abs_linear}a presents a colormap of the SLG absorbance under low--power excitation ($T_e=$ 300 K) as a function of the FF and the ribbons' filling factor, defined as $L = g / g_a$. The incident radiation excites hybrid electromagnetic modes that result from the strong coupling between the plasmonic resonance and the topological surface mode of the PCs. A detailed discussion of this interaction is available in Appendix \ref{AppA}. Figure~\ref{fig_abs_linear}b shows the SLG absorbance spectra for representative values of the filling factor $L$. Due to the negligible losses of the materials used to construct the PCs, all light absorption occurs within the SLG ribbons. The maximum absorbance is observed at $L=$~0.15--corresponding to a ribbon width of $g=1.2$ $\mu$m--where it reaches 50$\%$. This absorbance corresponds to the maximum achievable by a thin-film absorber in a symmetric two-port cavity, as dictated by the critical coupling condition \cite{piper2014total, doukas2018deep}. According to coupled mode theory \cite{piper2014total, xiao2020tailoring, doukas2018deep}, critical coupling occurs when, at the resonant frequency, the absorption rate of the absorber matches the decay rate of the cavity--formed in our case by the photonic crystals \cite{doukas2018deep, piper2014total}. For the critically coupled filling factor $L=0.15$, the hybrid resonance is centered at approximately 2.53 THz, slightly shifted from the surface state of the bare PCs, at 2.55 THz. This shift is expected because of the hybridization between the surface state mode and the localized plasmon resonance.

\begin{figure}[ht]\centering
	{\includegraphics[width = 0.9\textwidth]{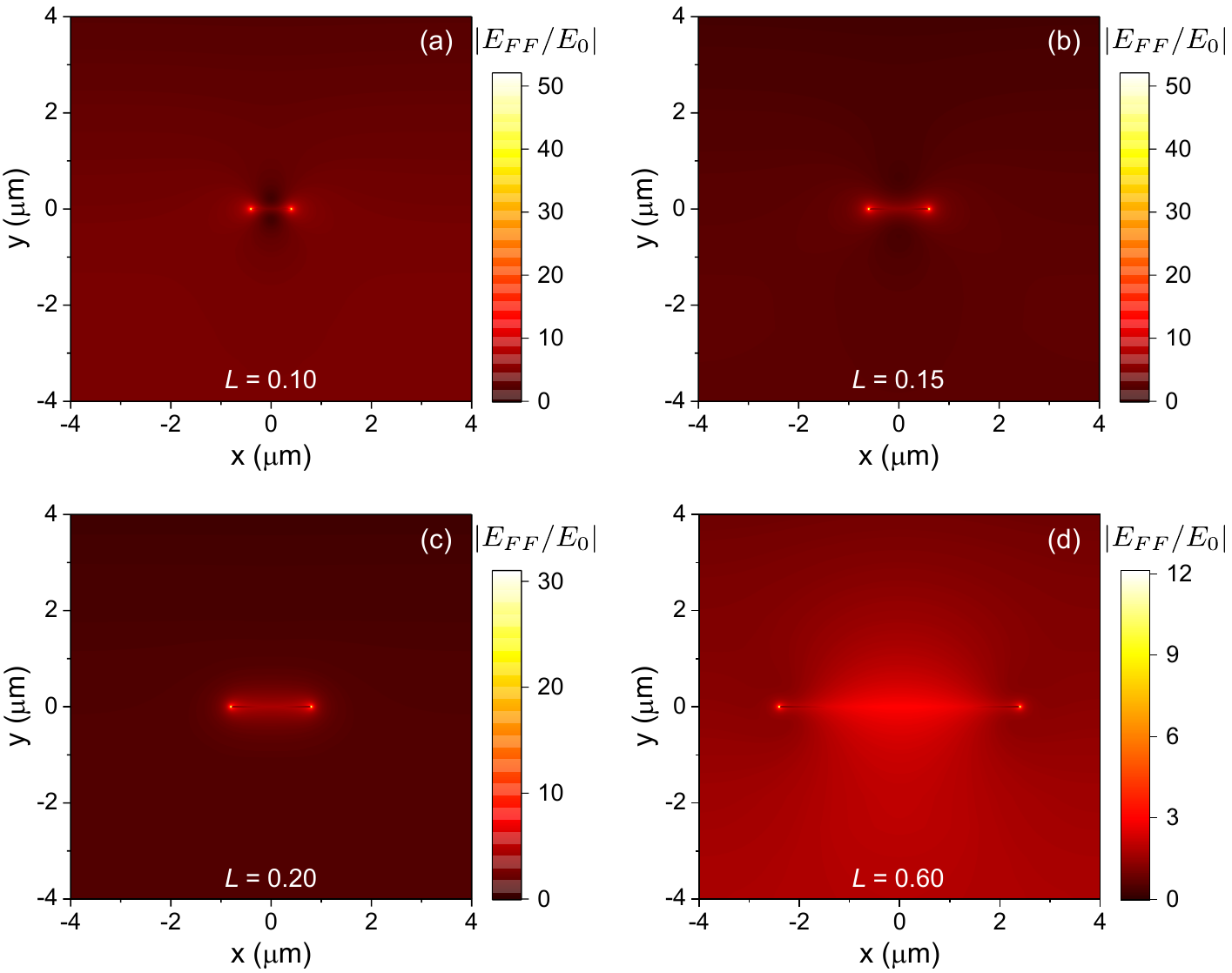}}
	\caption{Field enhancement assuming filling ratio (a) $L = 0.10$, (b) $L = 0.15$, (c) $L = 0.20$, (d) $L = 0.60$. All calculations were performed at the respected resonant frequency of each case, assuming low power excitation corresponding to $T_e$ = 300 K.}
	\label{field_enh}
\end{figure}
Figure~\ref{field_enh} shows the electric field (E--field) enhancement at the FF for the same $L$ values as in Fig.~\ref{fig_abs_linear}b. The frequency of maximum E--field enhancement coincides with that of peak absorption. For $L=0.15$, the E--field is strongly concentrated at the edges of the SLG ribbons--a signature of dipolar plasmonic resonances. In this optimal configuration, the plasmonic mode is resonant with the topological surface state, resulting in strong field confinement. As $L$ deviates from this value--for example, at $L=0.60$--the field enhancement weakens and becomes more spatially distributed, indicating a detuning from the hybrid resonance condition. Although $L=0.10$ also produces significant E--field enhancement, the conditions for critical coupling are not satisfied in this configuration, resulting in reduced SLG absorption. Specifically, at $L=0.15$, several mechanisms act synergistically to enhance THG. First, there is strong E--field enhancement at the FF, corresponding to the topological surface mode of the PCs. Since the PCs are also engineered to support a FP resonance at the TH frequency, higher--order plasmonic modes can be enhanced at this frequency as well. Simultaneously, the SLG absorbance is maximized, which leads to elevated $T_e$ and, consequently, an increased third--order conductivity $\sigma^{(3)}$.

\begin{figure}[ht]\centering
	{\includegraphics[width = 1.00\textwidth]{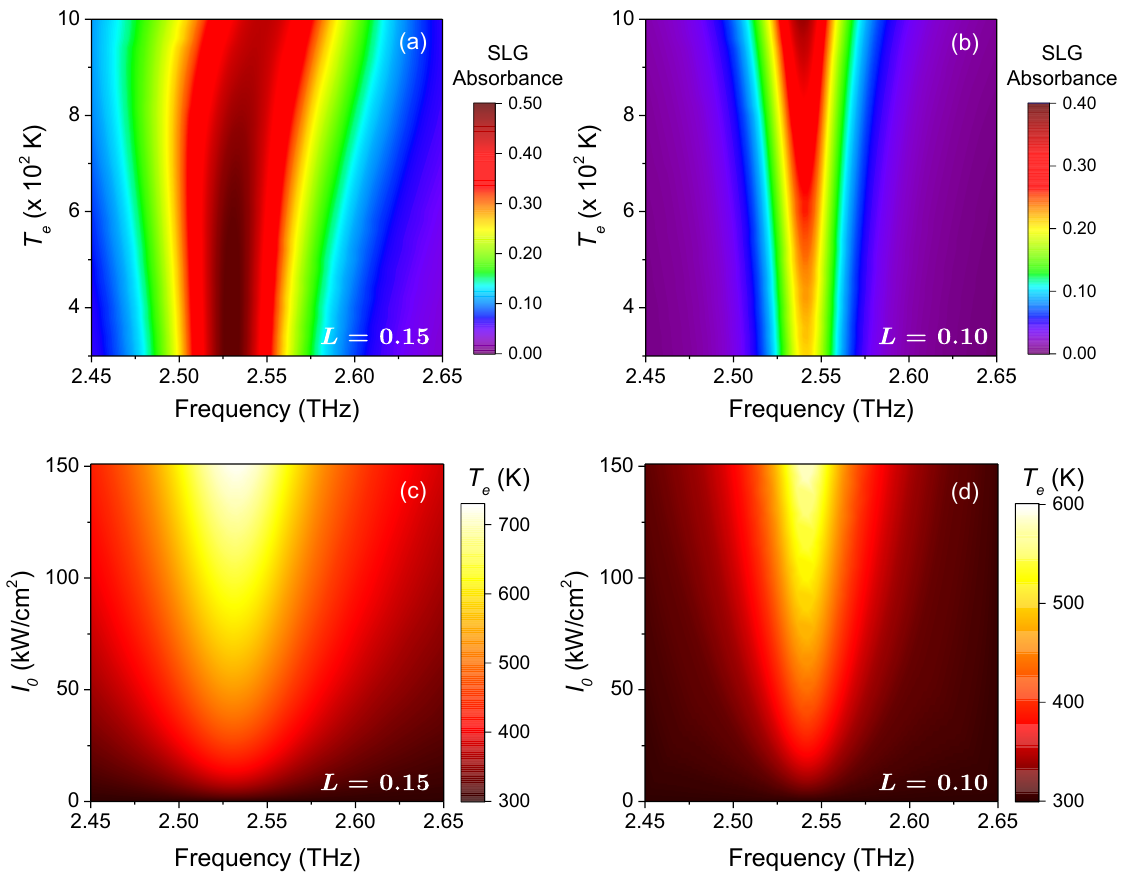}}
	\caption{SLG absorbance as a function of FF and $T_e$ for $L=$ 0.15 (a) and $L = 0.10$ (b). Photo--induced $T_e$ as a function of FF and $I_0$ for  $L=$ 0.15 (c) and $L = 0.10$ (d).}
	\label{fig_Te}
\end{figure}

Excitation with higher intensity THz radiation leads to SLG carrier thermalization, which modulates and shifts the plasmonic response \cite{renwen2020thermal} through thermally induced changes in SLG linear conductivity. In Fig. \ref{fig_Te} we plot SLG absorbance as a function of excitation frequency and $T_e$. We present the results for the two extreme cases of $L=0.15$ (Fig. \ref{fig_Te}a) and $L=0.10$ (Fig. \ref{fig_Te}b), which present the higher and lower absorption at $T_e= 300$ K respectively. The carrier temperature $T_e$ was arbitrarily set in the 300 K -- 1000 K range solely for the calculations shown in Fig.~\ref{fig_Te}a Fig.~\ref{fig_Te}b. In the former case ($L$ = 0.15), carrier temperatures beyond 500 K and the associated changes in SLG linear conductivity, lead to reduced absorbance. This is due to the disruption of critical coupling conditions, and is accompanied by a resonance shift to higher frequencies, a broader spectral response and a slightly reduced E--field enhancement at the FF. In contrast, for $L=0.10$, Fig. \ref{fig_Te}b shows that elevated carrier temperature drives the system closer to critical coupling, resulting in increased SLG absorbance. This is accompanied by an increased E--field enhancement at elevated temperatures. Using Eq. \ref{Te_cw}, as discussed in the methodology section, Fig.~\ref{fig_Te}c and Fig.~\ref{fig_Te}d presents the resulting electron temperature $T_e$ under THz photoexcitation as a function of excitation frequency and power density $I_0$. Results are once again shown for the two limiting cases from Fig. \ref{fig_abs_linear}b: $L=0.15$, with maximum initial (i.e., at $T_e$= 300 K) absorbance of 50$\%$ (Fig. \ref{fig_Te}c), and $L=0.10$, with an initial absorbance of 20$\%$ (Fig. \ref{fig_Te}d). Even at moderate intensities--power densities around $I_0=$  150 kW/cm$^2$, corresponding to an incident electric field amplitude of $E_0 \sim$ 10.6 kV/cm--carrier thermalization is substantial, with $T_e$ reaching approximately 750 K. These findings underscore the importance of incorporating the temperature--dependent calculations in the analysis carried out in this work.

\begin{figure}[ht]\centering
	{\includegraphics[width = 1.0\textwidth]{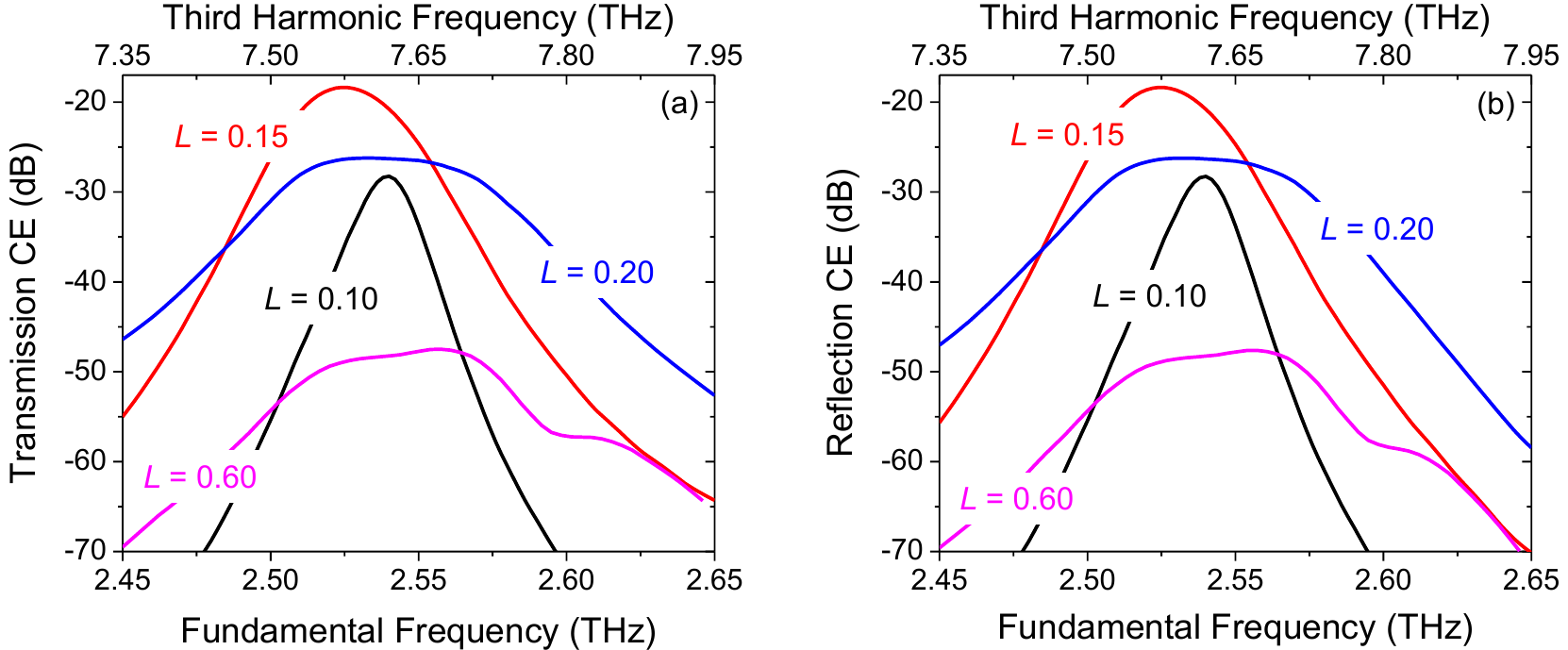}}
	\caption{Conversion efficiency spectra for transmitted (a) and reflected (b) third--harmonic  waves at different filling factors $L$. Excitation at the fundamental frequency (bottom horizontal axis) is assumed with a power density of $I_0=$ 150 kW/cm$^2$. The resulting TH waves are radiated at the corresponding third--harmonic frequencies (top horizontal axis).}
	\label{fig_CE}
\end{figure}

We move on with the calculations regarding the THG process. Figure \ref{fig_CE} plots the CE spectra for transmitted (Fig. \ref{fig_CE}a) and reflected (Fig. \ref{fig_CE}) TH waves for different $L$ values. The simulations were performed assuming excitation at the the FF with $I_0$ = 150 kW/cm$^2$, self--consistently accounting for $T_e$ increase and its effect in SLG response. We observe that the TH response is identical in both the forward (transmitted) and backward (reflected) directions. In fact, the equal CE for forward and backward TH waves holds across the entire range of input intensities considered in this work (see Appendix \ref{AppB}).  This symmetry arises from the TH waves being generated at the interface of the two adjacent PCs and coupling into a FP--like mode near $\sim $7.65 THz. The sharp transmission peak observed in the composite PC structure (Fig. \ref{fig_bandstructure}c) originates from this FP--like resonance at this frequency, which is supported by partial reflections at the boundaries of the two PCs. The periodic index modulation provides sufficient impedance contrast to sustain a standing--wave resonance across the full structure. The equal transmittance in both directions is a direct consequence of the identical band--structure of both PCs (Fig. \ref{fig_bandstructure}). 

Moreover, we find that even for filling factor values that shift the plasmonic resonance away from the FF of 2.55 THz--and consequently away from the FP mode at $\approx$~7.65 THz--the THG response can still exhibit a pronounced peak near 7.65 THz. For example, at $L=0.6$, a strong THG signal is observed in the vicinity of this TH frequency, despite the linear optical response, including SLG absorbance (Fig.~\ref{fig_abs_linear}) and the corresponding FF field enhancement, remaining nearly symmetric around the resonant FF which, in this case, corresponds to TH frequency at 7.8 THz. The apparent asymmetry in the THG spectrum is attributed to an increased local density of optical states (LDOS) near the FP resonance of the PCs. This Purcell--like effect enhances the radiative coupling efficiency of the induced TH radiation, resulting in stronger TH emission. Such cavity--enhanced emission is well established in both linear and nonlinear regimes and arises from the resonant amplification of available optical states \cite{celebrano2015mode}. Consequently, the configuration with $L=0.15$ meets both requirements, that is maximum field enhancement at FF and TH resonance in the vicinity of the FP mode at $\approx$~7.65 THz.
\begin{figure}[ht]\centering
	{\includegraphics[width = 1.0\textwidth]{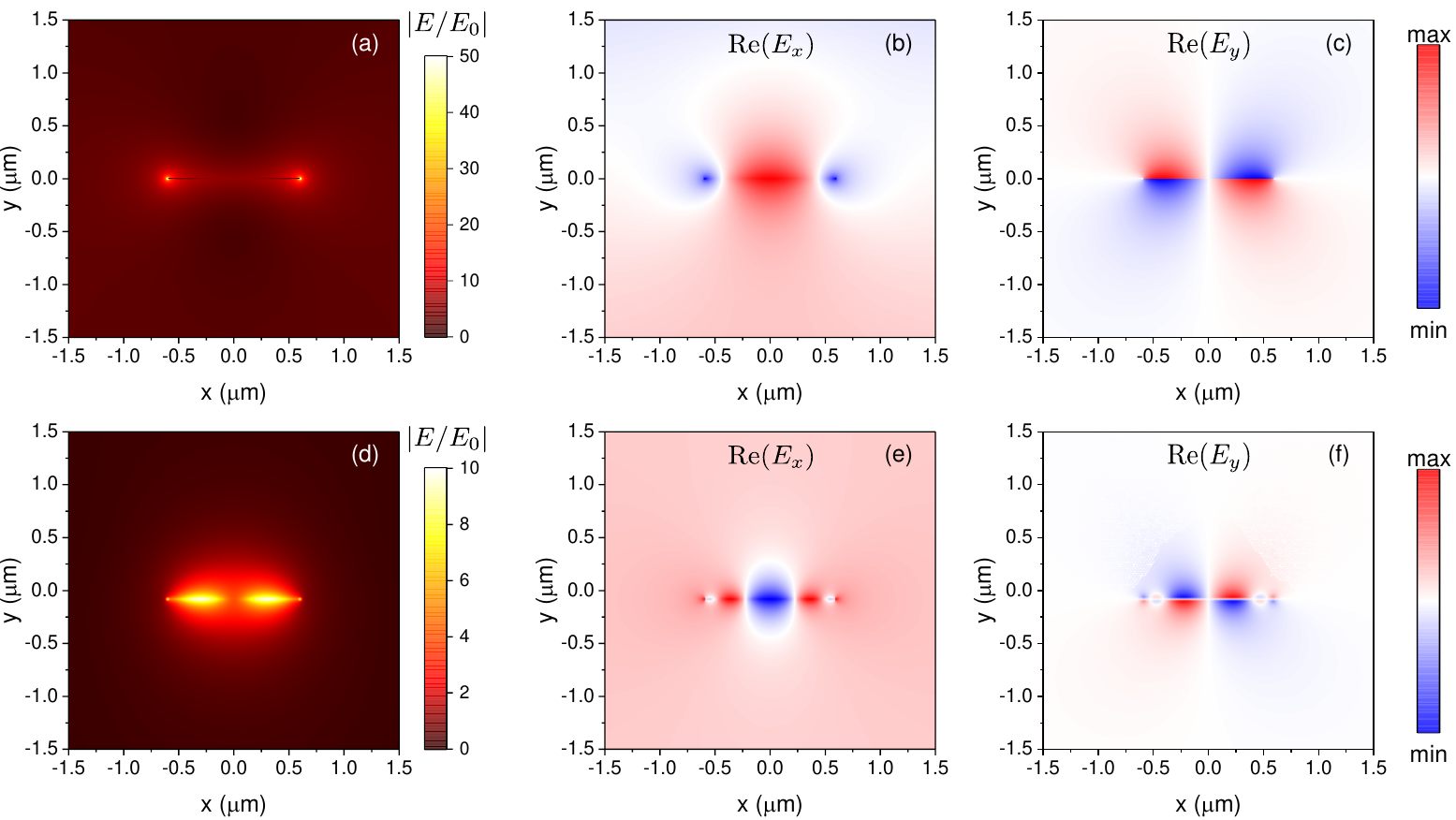}}
	\caption{Spatial distribution of electric field enhancement (a), real part of the x--component (b) and real part of the y--component for at FF 2.53 THz. Spatial distribution of electric field enhancement (d), real part of the x--component (e) and real part of the y--component (f) at TH frequency 7.59 THz. In all calculations $L = 0.15$ and excitation with $I_0=$ 150 kW/cm$^2$ was assumed.}
	\label{fields_f_015}
\end{figure}

Figure \ref{fields_f_015} plots the E--field enhancement spatial distribution both at the FF of 2.53 THz (top panel) and at the TH frequency of 7.59 THz (bottom panel), assuming $L=$ 0.15 and $I_0$ = 150 kW/cm$^2$. At the FF, the hybridization of the plasmonic resonance with the surface--mode results in a 50--fold enhancement of the incoming E--field (Fig. \ref{fields_f_015}a), while the distribution of individual field components (Figs. \ref{fields_f_015}b, \ref{fields_f_015}c) indicate a dipole plasmonic resonance. This configuration also leads to enhanced--light matter interaction at the TH frequency, due to the emergence of higher order plasmonic mode that resonates with the FP mode, resulting in a 10--fold E--field enhancement at the TH frequency (Fig. \ref{fields_f_015}d - Fig. \ref{fields_f_015}f).

\begin{figure}[ht]\centering
	{\includegraphics[width = 1.0\textwidth]{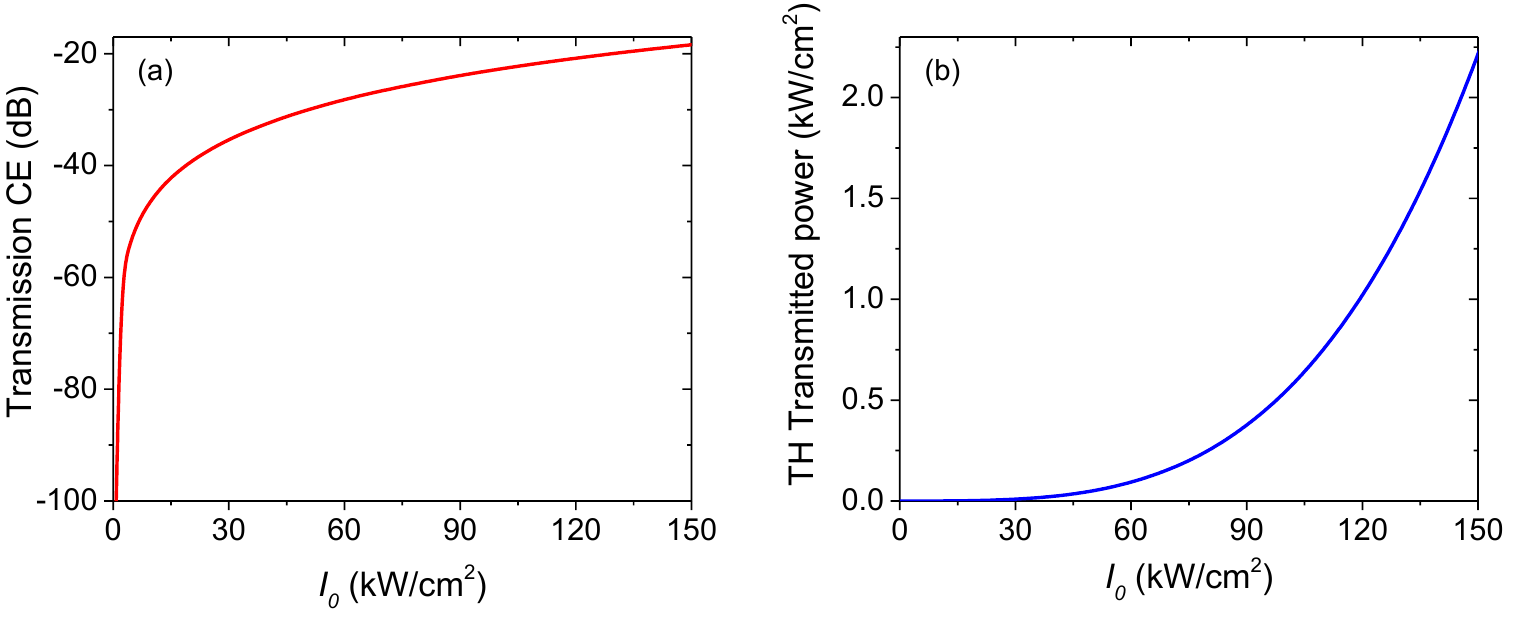}}
	\caption{Conversion efficiency (a) and transmitted power (b) as a function of $I_0$ assuming $L=0.15$ and excitation at 2.53 THz.}
	\label{fig_TH}
\end{figure}
Figure \ref{fig_TH} plots the transmission CE (panel a) and the transmitted TH power (panel b) for $L=$ 0.15 as a function of excitation power density $I_0$, assuming FF excitation at 2.53 THz, which corresponds to a TH frequency of 7.59 THz. The corresponding reflection CE and reflected TH power results are nearly identical to the results presented in Fig. \ref{fig_TH}. In both forward and backward directions, the CE increases with the incident intensity $I_0$, reaching $-$18.4 dB at $I_0=$ 150 kW/cm$^2$. The TH power outflow Fig. \ref{fig_TH}b presents the characteristic cubic $I_0$ dependence of the THG process \cite{jin2017enhanced, theodosi20212d}, surpassing 2 kW/cm$^2$ TH power for input intensity of $I_0 = $ 150 kW/cm$^2$.

\begin{figure}[ht]\centering
	{\includegraphics[width = 0.55\textwidth]{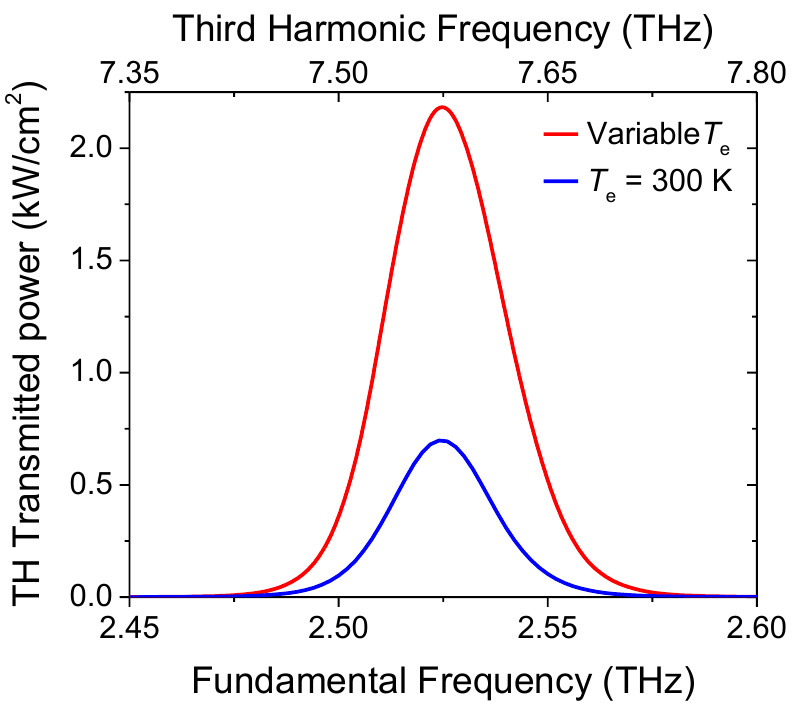}}
	\caption{TH transmitted power spectra assuming filling factor $L = 0.15$ and excitation intensity $I_0$ = 150 kW/cm$^2$. The red line corresponds to calculations self--consistently including $T_e$ increase effects. The blue line corresponds to calculations for constant $T_e$ = 300 K.}
	\label{fig_Te_THG}
\end{figure}

Finally, we focus on the impact of carrier heating and redistribution on the overall THG process. To illustrate this, Fig. \ref{fig_Te_THG} shows the transmitted power spectra for the optimal configuration with a filling factor $L=$ 0.15, assuming excitation with $I_0=$ 150 kW/cm$^2$, using two distinct approaches. The red line represents the results that include the effects of increased $T_e$ upon THz photoexcitation, as discussed in our previous analysis. On the same graph, we also plot the transmitted TH power assuming constant SLG $T_e=$ 300 K (blue line). In the first case, the transmitted power of TH waves is calculated to be $\approx$~2.2 kW/cm$^2$, yielding a CE of $-$18.4 dB. In contrast, neglecting the increased $T_e$  effects results in a transmitted power of approximately 0.7 kW/cm$^2$ corresponding to a CE of $-$23.3 dB. This difference arises from the modulation of graphene's nonlinear conductivity due to carrier heating. Although the increase in $\sigma^{(3)}$ with electron temperature $T_e$ is modest, the nonlinear nature of THG leads in a three--fold enhancement in the transmitted THG power. This comparison highlights the importance of including finite $T_e$ effects in accurate and effective design and modeling of THz THG.

\section{Conclusion}
In this study, we computationally investigated THz THG in SLG ribbons embedded at the interface of two PCs with inverse topological properties. The PC structure was designed to support a topological surface state at the FF and a FP resonance at the TH frequency. The width and periodicity of the SLG ribbons were chosen to support plasmonic resonances that strongly couple with both electromagnetic modes, enabling substantial local field enhancement at both FF and TH frequencies. Additionally, the configuration was optimized to maximize THz absorption of the incident radiation, thereby enhancing carrier thermalization. To accurately capture the temperature dependent nonlinear optical response of SLG under THz photoexcitation, we employed a self--consistent simulation framework that accounts for carrier heating. To the best of our knowledge, these results are the first to demonstrate that elevated carrier temperatures can be effectively harnessed to enhance third--harmonic radiation extraction in the THz regime.

Our results indicate that third--harmonic conversion efficiencies greater than 1 $\%$ ($-$20 dB) can be achieved at moderate input intensities ($I_0$ = 150 kW/cm$^2$), corresponding to a field strength of $E_0$ = 10.6 kV/cm. Notably, this performance is obtained without the need for electrostatic gating, thereby avoiding the associated fabrication complexities in periodic ribbon architectures. Furthermore, as the proposed configuration does not require a back mirror, third--harmonic waves can be extracted simultaneously in both forward (transmission) and backward (reflection) directions.

\appendix
 \section{Hybridization of graphene's plasmons in the topological photonic cavity}\label{AppA}

\begin{figure}[ht]\centering
		{\includegraphics{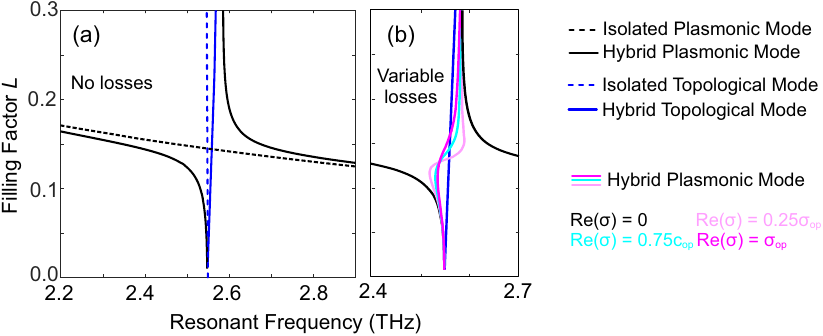}}
        \caption{Resonant frequencies of the isolated and hybrid dipolar plasmonic and topological mode in the  PC/graphene ribbons/PC*  system assuming (a) without losses and (b) with the presence of variable losses as a function of the graphene ribbons' filling factor in the structure. (a) Dashed lines correspond to the isolated systems, blue for the photonic cavity and black for the graphene plasmon. (b) Hybrid plasmonic/photonic modes with ascending losses.}
        \label{modes}
     \end{figure}

The PC/graphene--ribbon/PC* system supports both plasmonic and photonic modes, emerging at the graphene--ribbons and the interface of the PC/PC* photonic crystal topological cavity, respectively. The localized surface plasmons interact with the photonic cavity modes, leading to plasmon--photon hybridization. For the investigation of the plasmon--photon hybridization, we analyze the eigenmodes that the system supports, in the cases of the isolated (uncoupled) and hybrid (coupled) systems. We focus on the impact of graphene filling factor on the electromagnetic resonant characteristics of the system. The results are plotted in Fig.~\ref{modes}. Initially, we assume that there are no losses in the system, that is, for graphene ribbons, we assume $E_F=0.2$~ eV and we omit the lossy part of its conductivity. The  eigenfrequencies of the lossless case, with respect to the graphene filling factor, are present in Fig.~\ref{modes}(a). In particular, in Fig.~\ref{modes}(a) we plot, with dashed lines, the eigenfrequencies of the isolated systems. With a dashed blue line we plot the eigenfrequency of the PC/PC* topological cavity (no graphene ribbons) and with a dashed black line we plot the dipolar plasmonic mode of the ribbons, assuming a surrounding environment of refractive index equal to the median of $n_A$ and $n_B$, with respect to the filling factor of the ribbons, which varies in the range [0--0.3]. The photonic topological mode lies at $f$=2.55~THz. The dipolar plasmonic mode of the graphene ribbons presents a dispersion with respect to the filling factor and matches the topological cavity frequency, when the filling factor is around $L=$ 0.15. The eigenfrequencies of the coupled system, corresponding to the hybrid topological and plasmonic modes, are plotted in solid blue and solid black curves, respectively. The topological cavity mode experiences only a small shift in the presence of graphene ribbons, which increases as the ribbon gets wider. On the other hand, we observe that the dipolar plasmonic mode splits in to a lower and an upper branch as a result of the strong coupling between the plasmons in graphene and photons in the topological cavity. Figure~\ref{modes}(b) presents the evolution of the hybrid plasmonic/photonic modes with respect to the filling factor and losses, assuming ascending losses, i.e., $Re(\sigma)=0$, $Re(\sigma)=0.25\sigma_{max}$, $Re(\sigma)=0.75=\sigma_{max}$ and $Re(\sigma)=\sigma_{max}$. As the losses in graphene plasmons increase, the coupling between the two modes becomes less efficient and the system transitions from strong to weak coupling, leading to the deterioration of the  mode splitting. 
		 
\section{Frequency--Intensity dependence of THG process}\label{AppB}
		 
\begin{figure}[ht]\centering
{\includegraphics[width = 1.0\textwidth]{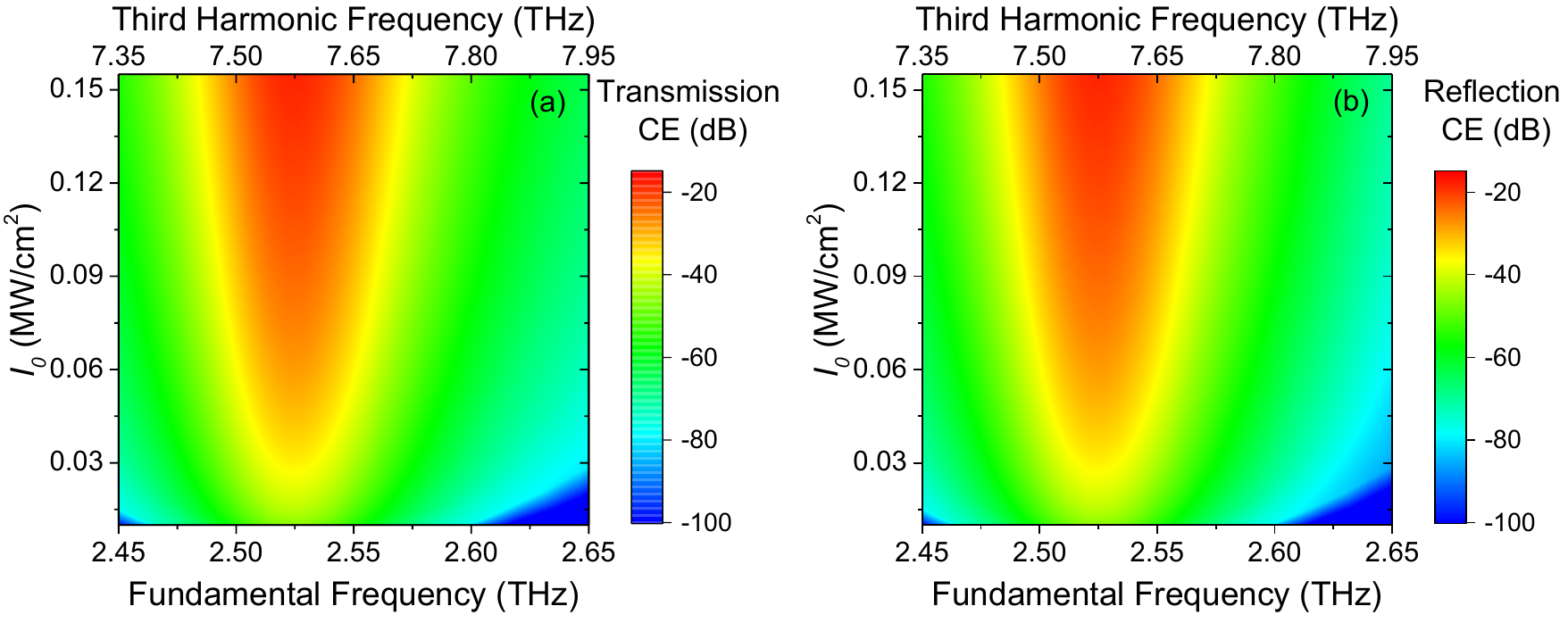}}
\caption{Transmitted (a) and reflected (b) conversion efficiency colormaps of the THG process as a function of excitation frequency and incident power density $I_0$. Bottom and top horizontal axis correspond to excitation  (fundamental) and generated third--harmonic frequencies respectively. }
\label{fig_CE_maps}
\end{figure}

For the optimal configuration with $L=0.15$, we present the CE colormaps for transmitted (Fig. \ref{fig_CE_maps}a) and reflected (Fig. \ref{fig_CE_maps}b) TH waves, as a function of excitation frequency and incident power density. Across the entire $I_0$  range, we observe nearly identical THG efficiency for both forward-- and backward--propagating TH radiation.
		 
\begin{acknowledgments}
This research work was supported by the National Recovery and Resilience
Plan Greece 2.0, funded by the European Union -- NextGenerationEU
(Implementation body: HFRI), Project Number: 14830, PhoToCon. 
\end{acknowledgments}

\bibliography{refs_Doukas}

\end{document}